# Interference Channels with Strong Secrecy

Xiang He  Aylin Yener

Wireless Communications and Networking Laboratory
Electrical Engineering Department
The Pennsylvania State University, University Park, PA 16802
xxh119@psu.edu  yener@ee.psu.edu

*Abstract*—It is known that given the real sum of two independent uniformly distributed lattice points from the same nested lattice codebook, the eavesdropper can obtain at most 1 bit of information per channel regarding the value of one of the lattice points. In this work, we study the effect of this 1 bit information on the equivocation expressed in three commonly used information theoretic measures, i.e., the Shannon entropy, the Rényi entropy and the min entropy. We then demonstrate its applications in an interference channel with a confidential message. In our previous work, we showed that nested lattice codes can outperform Gaussian codes for this channel when the achieved rate is measured with the weak secrecy notion. Here, with the Rényi entropy and the min entropy measure, we prove that the same secure degree of freedom is achievable with the *strong* secrecy notion as well. A major benefit of the new coding scheme is that the strong secrecy is generated from a single lattice point instead of a sequence of lattice points. Hence the mutual information between the confidential message and the observation of the eavesdropper decreases much faster with the number of channel uses than previously known strong secrecy coding methods for nested lattice codes.

## I. INTRODUCTION

Information theoretic secrecy, first proposed by Shannon [1], is an approach to study the secrecy aspect of a communication system against a *computation power unbounded* adversary. This approach was later applied to the wiretap channel [2]–[4] and recently extended to multiple access channel [5], broadcast channel [6], interference channel [7]. The focus of this body of literature is on the fundamental rate limits at which secret communication can take place.

Lattice codes were found recently to be useful in constructing information theoretically secure coding schemes [8]. In [8], [9], the authors showed that with the nested lattice code, the real sum of two lattice points leaked at most 1 bit of information per channel use to the eavesdropper regarding the value of one of the lattice points. Using this property, secrecy rate can then be achieved by using a random wiretap code while restricting the channel inputs to lattice points. The coding scheme is shown to be useful in providing secrecy in a multi-hop relay channel [8] and interference channels [9]–[12] . In particular, it outperforms Gaussian signaling scheme at high SNR in many fully connected two user Gaussian Channels with interference [9], [12].

The secrecy rates in all these works are derived with the notion of *weak* secrecy. If the observation of the eavesdropper is $Z^n$, and the confidential message is $W$, then weak secrecy requires:

$$\lim_{n\to\infty} \frac{1}{n} I(W; Z^n) = 0 \qquad (1)$$

However, sometimes, it is more useful to upper bound the following term:

$$\sum_{W,Z^n} |p(W, Z^n) - p(W) p(Z^n)| \qquad (2)$$

which tells how different the joint distribution of $\{W, Z^n\}$ is from this distribution if they are truly independent. Such a bound can be found via the Pinsker's inequality [13, Theorem 2.33] if $I(W; Z^n)$ can be bounded. Yet, it is clear that it is not possible to upper bound $I(W; Z^n)$ with the weak secrecy notion in (1). In fact, $I(W; Z^n)$ can still be arbitrarily large despite (1) being valid. For example, it can increase at the rate of $\log(n)$. To solve this problem, it is necessary to switch to the *strong* secrecy notion, which is defined as:

$$\lim_{n\to\infty} I(W; Z^n) = 0 \qquad (3)$$

In this work, we focus on how to achieve the strong secrecy notion using the nested lattice codes.

There are two known techniques in achieving strong secrecy. Reference [14] shows that for any discrete memoryless wiretap channel, it is possible for both the decoding error probability and $I(W; Z^n)$ to decrease exponentially with the number of the channel use. Yet, the proof of [14] uses random codes, and does not naturally extend to the lattice codes. One way to get around this problem is to treat each lattice point as a single channel use [15], [16], and design random codes for this extended channel using the method from [14]. However, as we will show later, doing so makes the length of a codeword to be much larger than the dimension of a lattice point. Because of this reason, the exponentially decreasing property of $I(W; Z^n)$ in $n$ is lost.

Reference [17] proposed that for a discrete memoryless channel, if there is a coding scheme that achieves weak secrecy given by (1), it can be combined with the privacy amplification technique in [18] to achieve the strong secrecy notion in (3). The exponential decreasing property of $I(W; Z^n)$ was not shown in [17] but can be proved by replacing the weak typicality used in [17] with strong typicality. Yet, since the proof of [17] relies on the typicality property of a i.i.d. generated sequence , it is again for random codes. If this approach were to be used with lattice codes, we will encounter the same problem we had with [14], i.e., the exponentially decreasing property of $I(W; Z^n)$ will be lost.

The coding scheme proposed in this work avoids this problem through a different method of lower bounding Rényi entropy and min entropy. In [17], this is obtained via the typicality property of i.i.d. sequence. Here it is proved instead using the representation theorem proposed in [8], [9], a unique property of nested lattice structure. The strong secrecy is hence generated from a single lattice point rather than a sequence of lattice points. Consequently, the exponentially decreasing property of $I(W; Z^n)$ is preserved.

The rest part of the paper is organized as follows. In Section II we summarize useful results on the effect of side information on information theoretic measures. These results show that revealing $m$ bits of information will not decrease entropy measures by more than $m$ with high probability. In Section III, we apply these results to lattice nested code, by viewing the integer in the representation theorem as the side information. We then use these results to prove achievable secure degree of freedom for an interference channel with confidential messages. The channel model is described in Section IV. The achievable secure degree of freedom is stated in Section V and proved in Section VI and Section VII. Section VIII concludes the paper.

## II. EFFECT OF SIDE INFORMATION ON INFORMATION THEORETIC MEASURE

In this section, we provide some supporting results which will be used later. Let $X, T$ denote two discrete random variable.

*Definition 1:* For a discrete random variable $X$, the Shannon entropy $H(X)$ is defined as

$$H(X) = -\sum_x \Pr(X=x) \log_2 \Pr(X=x) \quad (4)$$

The Rényi entropy $H_2(X)$ is defined as

$$H_2(X) = -\log_2 \sum_x \Pr(X=x)^2 \quad (5)$$

The min entropy $H_\infty(X)$ is defined as

$$H_\infty(X) = -\log_2 \max_x \Pr(X=x) \quad (6)$$

Let $\|T\|$ denote the cardinality of the alphabet set $T$ is defined on. Then we have the following lemmas.

*Lemma 1:* [19] For Shannon entropy, we have:

$$H(X|T) \geq H(X) - \log_2 \|T\| \quad (7)$$

The relationship says the introduction of $\log_2 \|T\|$ bits side information can not decrease the entropy of $X$ by more than $\log_2 \|T\|$ bits. The proof follows from the following equation:

$$H(X) \leq H(X, T) \quad (8)$$
$$= H(X|T) + H(T) \quad (9)$$
$$\leq H(X|T) + \log_2 \|T\| \quad (10)$$

In our previous work [8], [9], we use this lemma to prove that using nested lattice codes, the real sum of two lattice points from the same lattice codebook leaks *at most 1 bit of information per channel use* to the eavesdropper regarding the value of one of the lattice points, if the other point is independently generated and uniformly distributed over the codebook. This result opens the door for applying nested lattice codes to achieve weak secrecy in Gaussian channels.

*Lemma 2:* [20, P 106, Theorem 5.2] [17, Lemma 3] For Rényi entropy and $s > 0$, we have:

$$\Pr(t : H_2(X) - H_2(X|T=t) \leq \log_2 \|T\| + s)$$
$$\geq 1 - 2^{-(s/2-1)} \quad (11)$$

The lemma says with high probability the introduction of $\log_2 \|T\|$ bits side information can not decrease the Rényi entropy of $X$ by more than $\log_2 \|T\|$ bits. The proof can be found in [20, P 106, Theorem 5.2]. In [20], $T$ is also called "spoiler information". Later, we will use this lemma to prove strong secrecy rate based on the universal hash function.

*Lemma 3:* [17, Lemma 10] For min entropy and $s > 0$, we have:

$$\Pr(t : H_\infty(X) - H_\infty(X|T=t) \leq \log_2 \|T\| + s)$$
$$\geq 1 - 2^{-s} \quad (12)$$

The lemma says with high probability the introduction of $\log_2 \|T\|$ bits side information can not decrease the min entropy of $X$ by more than $\log_2 \|T\|$ bits. A proof of this lemma is provided in Appendix A. Later, we will use this lemma to prove strong secret key rate based on extractor functions [21].

## III. NESTED LATTICE CODES

In this section, we describe the nested lattice codes, the representation theorem from [8], [9] and its implication in terms of the three information theoretic measures described in Section II.

A nested lattice code is defined as an intersection of an $N$-dimensional "fine" lattice $\Lambda$ and the fundamental region of an $N$-dimensional "coarse" lattice $\Lambda_c$, denoted by $\mathcal{V}(\Lambda_c)$. $\Lambda, \Lambda_c \subset \mathbf{R}^N$. The term "nested" comes from the fact that $\Lambda_c \subset \Lambda$. The modulus operation is defined as the quantization error of a point $x$ with respect to the coarse lattice $\Lambda_c$:

$$x \bmod \Lambda_c = x - arg \min_{u \in \Lambda_c} \|x - u\|_2 \quad (13)$$

where $\|x - y\|_2$ is the Euclidean distance between $x$ and $y$ in $\mathbf{R}^N$. It can be verified that $\Lambda \cap \mathcal{V}(\Lambda_c)$ is a finite Abelian group when the addition operation between two elements $x, y \in \Lambda \cap \mathcal{V}(\Lambda_c)$ is defined as

$$x + y \bmod \Lambda_c \quad (14)$$

The signal $X^N$ transmitted over $N$ channel uses from a nested lattice codebook is given by

$$X^N = (u^N + d^N) \bmod \Lambda_c \quad (15)$$

Here $u^N$ is the lattice point chosen from $\Lambda \cap \mathcal{V}(\Lambda_c)$, and $d^N$ is called the dithering vector. Conventionally, $d^N$ is defined as a continuous random vector which is uniformly distributed over $\mathcal{V}(\Lambda_c)$ [22]. It was shown in [9] that a fixed dithering vector can be used. Either way, the nature of $d^N$ will not affect the result described below. In the following, we assume $u^N$ is independent from $d^N$. We also assume that $d^N$ is perfectly known by all receiving nodes, and hence, is not used to enhance secrecy.

As will be shown later, our goal in general will be to bound the expression of the following form: For Shannon entropy, it is written as:

$$H(u_1^N|X_1^N \pm X_2^N, d_1^N, d_2^N) \tag{16}$$

For Rényi entropy and min entropy, simply replace $H$ in (16) with $H_2$ and $H_\infty$ respectively. Here Here $u_i^N, X_i^N, d_i^N$ correspond to the $u^N, X^N, d^N$ mentioned above respectively. That is to say that $u_i^N \in \Lambda \cap \mathcal{V}(\Lambda_c)$; $d_i^N$ is the dithering noise; $X_i^N = (u_i^N + d_i^N) \mod \Lambda_c$. In addition, $u_i^N, d_i^N, i = 1, 2$ are independent.

All three information theoretic measures considered in this work can be bounded using the *representation theorem* from [8], [9]:

*Theorem 1:* Let $t_1, t_2, ..., t_K$ be $K$ numbers taken from the fundamental region of a given lattice $\Lambda$. There exists a integer $T$, such that $1 \leq T \leq K^N$, and $\sum_{k=1}^{K} t_k$ is uniquely determined by $\{T, \sum_{k=1}^{K} t_k \mod \Lambda\}$.

The proof can be found in [9].

Clearly, $X_i^N, i = 1, 2$ are in the fundamental region of $\Lambda_c$. Hence the theorem says $X_1^N + X_2^N$ can be uniquely determined by $T, (X_1^N + X_2^N) \mod \Lambda_c, 1 \leq T \leq 2^N$. Since fundamental region of $\Lambda_c$ is symmetric with respect to the origin, the same results hold for $X_1^N - X_2^N$ as well.

In the sequel, let $T$ be the integer whose existence is guaranteed by Theorem 1. For a discrete random variable $X$, let $\bar{x}$ denote a deterministic value such that $\Pr(X = \bar{x}) > 0$. With these notations, we have the following corollary to the three lemmas in Section II.

*Corollary 1:* In terms of Shannon entropy, we have

$$H\left(u_1^N|u_1^N \pm u_2^N \mod \Lambda_c, d_i^N, i = 1, 2\right) \\ - H\left(u_1^N|X_1^N \pm X_2^N, d_i^N, i = 1, 2\right) \leq \log_2 \|T\| \tag{17}$$

In terms of Rényi entropy, consider the inequality

$$H_2(u_1^N|u_1^N \pm u_2^N \mod \Lambda_c = \bar{u}, d_i^N = \bar{d}_i^N, i = 1, 2) \\ - H_2\left(u_1^N|X_1^N \pm X_2^N = \bar{x}, d_i^N = \bar{d}_i^N, i = 1, 2\right) \\ \leq \log_2 \|T\| + s \tag{18}$$

where $\bar{x}$ is a function of $\bar{u}, \bar{d}_i^N$ and $T$. Then we have

$$\Pr\left(\bar{t} : T = \bar{t}, \text{ and (18) holds }\right) \geq 1 - 2^{-(s/2-1)} \tag{19}$$

In terms of min entropy, consider the inequality:

$$H_\infty\left(u_1^N|u_1^N \pm u_2^N \mod \Lambda_c = \bar{u}, d_i^N = \bar{d}_i^N, i = 1, 2\right) \\ - H_\infty\left(u_1^N|X_1^N \pm X_2^N = \bar{x}, d_i^N = \bar{d}_i^N, i = 1, 2\right) \\ \leq \log_2 \|T\| + s \tag{20}$$

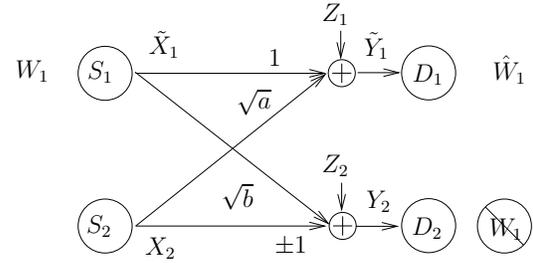

Fig. 1. The Gaussian Wiretap Channel with a Cooperative Jammer

we have

$$\Pr\left(\bar{t} : T = \bar{t}, \text{ and (20) holds }\right) \geq 1 - 2^{-s} \tag{21}$$

*Proof:* Equation (17) uses Lemma 1 and its proof can be found in [8], [9].

We next prove (19). We begin with:

$$H_2\left(u_1^N|X_1^N \pm X_2^N = \bar{x}, d_i^N = \bar{d}_i^N, i = 1, 2\right) \tag{22}$$

$$=H_2(u_1^N| \sum_{j=1}^{2} \left(u_j^N + d_j^N\right) \mod \Lambda_c = \bar{x}, \\ d_i^N = \bar{d}_i^N, i = 1, 2) \tag{23}$$

From Theorem 1, (23) can be written as:

$$H_2(u_1^N|(\sum_{j=1}^{2} u_j^N + d_j^N) \mod \Lambda_c = \bar{x}', \\ T = \bar{t}, d_i^N = \bar{d}_i^N, i = 1, 2) \tag{24}$$

where $\bar{x}'$ is a constant such that when $(\sum_{j=1}^{2} u_j^N + d_j^N) \mod \Lambda_c = \bar{x}'$ and $T = \bar{t}$, we have $\sum_{j=1}^{2} \left(u_j^N + d_j^N\right) \mod \Lambda_c = \bar{x}$. Theorem 1 guarantees the existence of such $\bar{x}'$ and $\bar{t}$.

We then consider $T$ as the side information and apply Lemma 2 to (24). This yields (19).

The proof for (21) is similar. We simply rewrite (23)-(24) by replacing $H_2$ with $H_\infty$. Equation (21) follows by viewing $T$ as the side information and apply Lemma 3.

Hence we have proved the Corollary. ∎

We next demonstrate the usefulness of Corollary 1 in an interference channel with confidential message.

IV. CHANNEL MODEL AND PROBLEM FORMULATION

We will focus on the Gaussian wiretap channel with a cooperative jammer [23]. All results derived in this work extend to the $K$-user interference channel with a confidential message in a straightforward manner.

The channel model is shown in Figure 1. As shown in this figure, after normalizing the channel gains of the two intended links to 1, the received signals at the two receiving node $D_1$ and $D_2$ can be expressed as

$$\tilde{Y}_1 = \tilde{X}_1 + \sqrt{a}X_2 + Z_1 \\ Y_2 = \sqrt{b}\tilde{X}_1 \pm X_2 + Z_2 \tag{25}$$

where $Z_i, i = 1, 2$ is a zero-mean Gaussian random variable with unit variance, and $\sqrt{a}$, $\sqrt{b}$ and $Z_i$ are real numbers.

As in [9], we let $X_1 = \sqrt{b}\tilde{X}_1$ and $Y_1 = \sqrt{b}\tilde{Y}_1$. Then from (25), we have

$$Y_1 = X_1 + \sqrt{ab}X_2 + \sqrt{b}Z_1 \qquad (26)$$
$$Y_2 = X_1 \pm X_2 + Z_2$$

In the sequel, we will focus on this scaled model which will be more convenient to explain our results.

In [9], we calculate the weak secrecy rate for this model. In this work, we will calculate the strong secrecy rate and strong secrecy key rate, which we shall define shortly.

In the strong secrecy rate problem, node $S_1$ sends a message $W_1$ via $\tilde{X}_1$ to node $D_1$, which must be kept secret from node $D_2$. Let $M_1$ be the local randomness at $S_1$ and $f_i$ be its encoding function at the $i$th channel use. Then we have:

$$X_{1,i} = f_i(W_1, M_1) \qquad (27)$$

Node $S_2$, the cooperative jammer, sends signal $X_2$.

Let $\hat{W}_1$ be the estimate of $W_1$ by node $D_1$. For $D_1$ to receive $W_1$ reliably, we require

$$\lim_{n\to\infty} \Pr\left(W_1 \neq \hat{W}_1\right) = 0 \qquad (28)$$

In addition, since $W_1$ must be kept secret from $D_2$, we require the strong secrecy notion as defined in (3), which takes the following expression for this model:

$$\lim_{n\to\infty} I(W_1; Y_2^n) = 0 \qquad (29)$$

The achieved secrecy rate $R_e$ is defined as:

$$R_e = \lim_{n\to\infty} \frac{1}{n} H(W_1) \qquad (30)$$

such that the conditions (28), (29) are fulfilled simultaneously.

In the secret key generation problem, node $S_1$ and $D_1$ communicate for $n$ channel uses and after that wants to generate the same key from the signals available to them. Let $g_1, g_2$ be the generation function used by $S_1$ and $D_1$ respectively. Let $M_1, M_2$ be their local randomness respectively. Then

$$K_1 = g_1(M_1) \qquad (31)$$
$$\hat{K}_1 = g_2(M_2, Y_1^n) \qquad (32)$$

The encoding function at node 1 for the $i$th channel use is defined as

$$X_{1,i} = f_i(M_1) \qquad (33)$$

and we require:

$$\lim_{n\to\infty} \Pr(K_1 \neq \hat{K}_1) = 0 \qquad (34)$$
$$\lim_{n\to\infty} I(K_1; Y_2^n) = 0 \qquad (35)$$

The achieved secrecy key rate $R_{k,e}$ is defined as

$$R_{k,e} = \lim_{n\to\infty} \frac{1}{n} H(K_1) \qquad (36)$$

such that the conditions (34), (35) are fulfilled simultaneously.

For both problems, there are two constraints on the input distribution to the channel model in (26): First, we assume there is no common randomness shared by the encoders of $S_1$ and $S_2$. This means, the input distribution to the channel is constrained to be

$$p(X_1^m) p(X_2^m) \qquad (37)$$

where $m$ is the number of channel uses involved. Intuitively, this implies that if $X_2$ is employed to send interference to confuse the eavesdropper, its effect can not be mitigated by coding $X_1$ via dirty-paper coding [24].

Second, the average power of $X_i$ is constrained to be $\bar{P}_i$. If $X_{i,j}$ is the $j$th component of $X_i$, this means:

$$\lim_{m\to\infty} \frac{1}{m} \sum_{j=1}^{m} E\left[|X_{i,j}|^2\right] \leq \bar{P}_i, i = 1, 2 \qquad (38)$$

In the next section, we examine the high SNR behavior of $R_e$ and $R_{e,k}$ when these two requirements on the channel input distribution are fulfilled.

## V. MAIN RESULTS

*Definition 2:* The secure degree of freedom of the secrecy rate is defined as:

$$\text{s.d.o.f.} = \limsup_{\bar{P}_i \to \infty, i=1,2} \frac{R_e}{\frac{1}{2}\log_2\left(\sum_{i=1}^{2} \bar{P}_i\right)} \qquad (39)$$

The secure degree of freedom of the secrecy key rate is defined in the same way by replacing $R_e$ in (39) with $R_{k,e}$.

We also notice that any $\sqrt{ab}$, $\sqrt{ab} \neq 0$, can be represented in the following form:

$$\sqrt{ab} = p/q + \gamma/q \qquad (40)$$

where $p, q$ are positive integers, and $-1 < \gamma < 1, \gamma \neq 0$. In this case, the channel model (26) can be expressed as:

$$qY_1 = qX_1 + (p+\gamma) X_2 + q\sqrt{b}Z_1 \qquad (41)$$
$$Y_2 = X_1 \pm X_2 + Z_2 \qquad (42)$$

Using this notation, we have the following theorem regarding the achievable secure degree of freedom:

*Theorem 2:* There exists a encoder $\{f_i\}$, such that the following secure degree of freedom is achievable using nested lattice codes when $0 < |\gamma| < 0.5$:

$$\left[\frac{0.25 \log_2(\alpha) - 1}{\frac{1}{2} \log_2(\alpha\beta + 1)}\right]^+ \qquad (43)$$

where

$$\alpha = \frac{1 - 2\gamma^2 + \sqrt{1 - 4\gamma^2}}{2\gamma^4} \qquad (44)$$

and

$$\beta = q^2 + (p+\gamma)^2 \qquad (45)$$

such that for any given transmission power

$$\lim_{n \to \infty} -\frac{1}{n} \log_2 I(W_1; Y_2^n) > 0 \quad (46)$$

where $n$ denotes the total number of channel uses.

*Theorem 3:* There are generation functions $g_j, j = 1, 2$ with explicit construction, such that (43) is the achievable secure degree of freedom for the key generation problem, and for any given transmission power:

$$\lim_{n \to \infty} -\frac{1}{\sqrt{n}} \log_2 I(K_1; Y_2^n) > 0 \quad (47)$$

where $n$ denotes the total number of channel uses.

*Remark 1:* In [9], we proved that (43) is the achievable s.d.o.f. for the weak secrecy notion. Here, we show that the same s.d.o.f. is achievable for the strong secrecy notion.

*Remark 2:* Reference [14], [17] show that using the strong secrecy notion instead of the weak one did not decrease the achievable secrecy rate. This coincides with the observation in Remark 1. Yet, the proofs in [14], [17] use random codes, and does not naturally extend to the lattice codes. One way to get around this problem is to treat each lattice point as a single channel use [15], and design random codes for this extended channel using the method [14]. However, as mentioned in the introduction, doing so will not achieve the exponential decrease property as given by (46) and (47). To see that, let each codeword in this randomly generated codebook contains $M$ lattice points, and the dimension of the lattice be $N$. Then it was proved in [22] that the decoding error probability decreases exponentially with respect to $N$. Also from the coding scheme in [14], we have $I(W; Y_2^n)$ decreases exponentially with respect to $M$. Since the total number of channel uses in this case is $MN$, (46) and (47) no longer holds unless $N$ does not increase proportionally with $M$. Hence, unless a significant price is paid on the decoding error probability, the exponential decrease in $I(W; Y_2^n)$ can not be preserved.

*Remark 3:* The proof of Theorem 2 uses universal hash function and is existential. The secrecy notion achieved by Theorem 3 is weaker but its proof is constructive. The proof uses on the extractor function proposed in [21], which gives explicit construction of the function.

## VI. Proof of Theorem 2

We first introduce some useful results on universal hash functions.

*Definition 3:* [18, Definition 1] A set of functions $\mathcal{A} \to \mathcal{B}$ is a class of *universal hash function* if for a function $g$ taken from the set according to a uniform distribution, and $x_1, x_2 \in \mathcal{A}, x_1 \neq x_2$, the probability that $g(x_1) = g(x_2)$ is at most $1/|\mathcal{B}|$.

*Theorem 4:* [18, Corollary 4] Let $G$ be selected according to a uniform distribution from a class of universal hash function from $A$ to $\mathbf{GF}(q)^r$. For two random variables $A, B$, if for a constant c, $H_2(A|B=b) > c$, then

$$H(G(A)|G, B=b) > r \log_2 q - \frac{2^{r \log_2 q - c}}{\ln 2} \quad (48)$$

Let $G$ be taken from a set of linear mapping from $\mathbf{GF}(q)^N$ to $\mathbf{GF}(q)^r$ according to a uniform distribution. Hence $G$ can be represented as a matrix over $\mathbf{GF}(q)$ with $r$ rows and $N$ columns. For this class of $G$, we have the following lemma:

*Lemma 4:* The probability that $G$ has full row rank is greater than $1 - q^{r-N}$.

*Proof:* Let $g_i, i = 1, ..., r$ be the $i$th row of $G$. Then $G$ does not have full row rank if and only if

$$a_1 g_1 + a_2 g_2 + ... + a_r g_r = 0, \quad a_i \in \mathbf{GF}(q) \quad (49)$$

Since at least one $a_i$ has to be non-zero, there are $q^r - 1$ possible choice for $a_i$.

For each choice of $\{a_i\}$, since one $a_i$ is not zero, there are $q^{N(r-1)}$ solutions for $\{g_i\}$. Hence there are at most $q^{N(r-1)}(q^r - 1)$ $G$s that do not have full row rank. There are $q^{Nr}$ possible $G$s in all, each chosen with equal probability. Hence the probability that $G$ does not have full row rank smaller than $q^{r-N}$, and we have Lemma 4. ∎

The reason that we are interested in this class of $G$ is because of the following lemma:

*Lemma 5:* [18] The set of linear mapping as defined in Lemma 4 is a class of universal hash function.

We then briefly describe the lattice codebook from [9], which we shall generate strong secrecy from.

In this coding scheme, node $S_k$ sends the signal $X_k^N$ over $N$ channel uses. $X_k^N$ is the sum of codewords from several layers as shown below:

$$X_k^N = \sum_{i=1}^{M} X_{k,i}^N, \quad k = 1, 2 \quad (50)$$

where $M$ is the total number of layers in use. $X_{k,i}^N$ is the signal sent by the $S_k$ in the $i$th layer. For each layer, we use the nested lattice code described in Section III. Let $\{\Lambda_i, \Lambda_{c,i}\}$ be nested lattice pair assigned to layer $i$. Then the signal $X_{k,i}^N$ is computed according to this nested lattice pair as:

$$X_{k,i}^N = \left(u_{k,i}^N + d_{k,i}^N\right) \mod \Lambda_{c,i} \quad k = 1, 2, i = 1, ..., M \quad (51)$$

where $d_{k,i}^N$ is the dithering vector, uniformly distributed over $\mathcal{V}(\Lambda_{c,i})$, perfectly known by all receiving nodes and independently generated for each node, each layer and each block of $N$ channel uses. Let $u_{k,i}^N$ be the lattice point such that:

$$u_{k,i}^N \in \mathcal{V}(\Lambda_{c,i}) \cap \Lambda_i, \quad k = 1, 2 \quad (52)$$

Note that both node $S_1$ and $S_2$ use the same lattice codebook for each layer. We choose the input distribution to the channel such that $u_{k,i}^N$ is uniformly distributed over $\mathcal{V}(\Lambda_{c,i}) \cap \Lambda_i$ and $u_{k,i}^N$ is independent between each node, each layer and each bloc of $N$ channel channel uses.

Define $R_i, i = 1, ..., M$ as the rate of the codebook for the $i$th layer:

$$R_i = \frac{1}{N} \log_2 |\mathcal{V}(\Lambda_{c,i}) \cap \Lambda_i| \quad (53)$$

Define $R_0$ as the average rate per layer:

$$R_0 = \frac{1}{M} \sum_{i=1}^{M} R_i \quad (54)$$

Define $\Lambda^M \cap \mathcal{V}(\Lambda_c)^M$ as the $M$-fold Cartesian product of $\Lambda_i \cap \mathcal{V}(\Lambda_{c,i}), i = 1...M$. Note that $\Lambda^M \cap \mathcal{V}(\Lambda_c)^M$ is also a nested lattice codebook. Its dimension is denoted by $\bar{N} = MN$. Define $t_i^{\bar{N}}, i = 1, 2$ as the Cartesian product of $u_{k,i}^N, k = 1,...,M, i = 1, 2$. Define $t_1^{\bar{N}} \oplus t_2^{\bar{N}}$ as $u_{k,1}^N + u_{k,2}^N \mod \Lambda_c, k = 1,...,M$. The shorthand $d = \bar{d}$ to denote $d_j^{\bar{N}} = \bar{d}_j^{\bar{N}}, j = 1, 2$.

Let $\lfloor x \rfloor$ be the operation that rounds $x$ to the nearest integer less than or equal to $x$. Define $\bar{N}_0$ as

$$\bar{N}_0 = \left\lfloor \log_2 |\Lambda^M \cap \mathcal{V}(\Lambda_c)^M| \right\rfloor \quad (55)$$

Then

$$\bar{N}_0 \geq \bar{N} R_0 - 1 \quad (56)$$

Choose the subset $K$ of the codebook $(\Lambda + d_1^N)^M \cap \mathcal{V}(\Lambda_c)^M$ that yields the minimal average decoding error probability with the lattice decoder and has size $|K| = 2^{\bar{N}_0}$. Define $v$ as the one-to-one mapping from $K$ to $\mathbf{GF}(2^{\bar{N}_0})$.

We begin with the fact that $t_1^{\bar{N}} \oplus t_2^{\bar{N}}$ is independent from $t_1^{\bar{N}}$, from which we have:

$$H_2\left(v(t_1^{\bar{N}})|t_1^{\bar{N}} \oplus t_2^{\bar{N}} = t^{\bar{N}}, d = \bar{d}\right) = H_2\left(v(t_1^{\bar{N}})\right) \quad (57)$$

According to Corollary 1, we have, for a given integer $a$, $1 \leq a \leq 2^{\bar{N}}$ and $t^{\bar{N}}$ taken from $\Lambda^M \cap \mathcal{V}(\Lambda_c)^M$, with a probability of at least $1 - 2^{-(s/2-1)}$:

$$H_2\left(v(t_1^{\bar{N}})|t_1^{\bar{N}} \oplus t_2^{\bar{N}} = t^{\bar{N}}, d = \bar{d}, T = a\right) \quad (58)$$
$$\geq H_2\left(v(t_1^{\bar{N}})|t_1^{\bar{N}} \oplus t_2^{\bar{N}} = t^{\bar{N}}, d = \bar{d}\right) - \log_2 |T| - s \quad (59)$$
$$= \bar{N}_0 - \bar{N} - s \quad (60)$$
$$\leq \bar{N}(R_0 - 1) - 1 - s \quad (61)$$

where (61) follows from (56).

For a positive integer $\bar{r}_0$, let $G$ be a random function which is uniformly distributed over the set of linear functions from $\mathbf{GF}(2)^{\bar{N}_0}$ to $\mathbf{GF}(2)^{\bar{r}_0}$. Then when (58)-(61) holds, we have the following equations because of Theorem 4:

$$H\left(G\left(v(t_1^{\bar{N}})\right)|G, t_1^{\bar{N}} \oplus t_2^{\bar{N}} = t^{\bar{N}}, d = \bar{d}, T = a\right) \quad (62)$$
$$\geq \bar{r}_0 - \frac{2^{\bar{r}_0 - c}}{\ln 2} \quad (63)$$

where $c = \bar{N}(R_0 - 1) - 1 - s$.

Since the probability that (58)-(61) holds is at least $1 - 2^{-(s/2-1)}$, we have:

$$H\left(G\left(v(t_1^{\bar{N}})\right)|G, t_1^{\bar{N}} \oplus t_2^{\bar{N}}, d, T\right) \quad (64)$$
$$\geq \left(1 - 2^{-(s/2-1)}\right)\left(\bar{r}_0 - \frac{2^{\bar{r}_0 - c}}{\ln 2}\right) \quad (65)$$

Choose $s = \varepsilon \bar{N}$, where $0 < \varepsilon < R_0 - 1$. Choose $\bar{r}_0$ such that for $\delta > 0$:

$$\bar{r}_0 < \bar{N}(R_0 - 1) - s - \bar{N}\delta \leq \bar{N}_0 - \bar{N} - s - \bar{N}\delta/2 \quad (66)$$

It can be verified for this $\bar{r}_0$ we have

$$0 \leq \bar{r}_0 \leq \bar{N}[R_0 - 1 - \varepsilon']^+ \quad (67)$$

for a constant $\varepsilon' > 0$. For this $\bar{r}_0$ and $s$, from (65), we observe that there exists $\beta > 0$, such that

$$I\left(G\left(v(t_1^{\bar{N}})\right); t_1^{\bar{N}} \oplus t_2^{\bar{N}}, d, T | G\right) \leq e^{-\beta \bar{N}} \quad (68)$$

We next use the fact that for sufficiently large $N$, hence $\bar{N}$, most $G$ has full row rank as shown in Lemma 4. Therefore, there must exists a $G = g$, such that
1) $g$ has full rank.
2) $I\left(G\left(v(t_1^{\bar{N}})\right); t_1^{\bar{N}} \oplus t_2^{\bar{N}}, T | G = g\right) \leq 2e^{-\beta \bar{N}}$

Note that this $g$ is chosen for a uniform distribution for $t_i^{\bar{N}}, i = 1, 2$, $t_1^{\bar{N}}$ and $t_2^{\bar{N}}$ being independent and is not necessarily a good choice for other channel input distribution.

This result can then be used to an encoder with rate arbitrarily close to $[M(R_0 - 1)]^+$, as shown below:

Let $g'$ be $(\bar{N}_0 - \bar{r}_0) \times \bar{N}_0$ matrix such that $\begin{bmatrix} g' \\ g \end{bmatrix}$ is invertible. Define $S$ such that $S'$ such that

$$\begin{bmatrix} g'_{(\bar{N}_0 - \bar{r}_0) \times \bar{N}_0} \\ g_{\bar{r}_0 \times \bar{N}_0} \end{bmatrix} v(t_1^{\bar{N}}) = \begin{bmatrix} S'_{(\bar{N}_0 - \bar{r}_0) \times 1} \\ S_{\bar{r}_0 \times 1} \end{bmatrix} \quad (69)$$

Then $S = g(v(t_1^{\bar{N}}))$. Define $A$ as the inverse of $\begin{bmatrix} g' \\ g \end{bmatrix}$, then the encoder is given by:

$$t_1^{\bar{N}} = v^{-1} A \begin{bmatrix} S'_{(\bar{N}_0 - \bar{r}_0) \times 1} \\ S_{\bar{r}_0 \times 1} \end{bmatrix} \quad (70)$$

where $S \in \mathbf{GF}(2^{\bar{r}_0})$ be the input to the encoder. We assume $S$ is uniformly distributed over $\mathbf{GF}(2^{\bar{r}_0})$. $t_1^{\bar{N}} \in \Lambda^M \cap \mathcal{V}(\Lambda_c)^M$ be its output. $S'$ represents the randomness in the encoding scheme. We observe that, if $\{S'_{(\bar{N}_0 - \bar{r}_0) \times 1}, S_{\bar{r}_0 \times 1}\}$ is uniformly distributed over $\mathbf{GF}(2)^{\bar{N}_0}$ and (70) is used as the encoder, $t_1^{\bar{N}}$ is also uniformly distributed over the set $K$. Since $G = g$ is chosen when $t_1^{\bar{N}}$ has a uniform distribution over $K$, this means when (70) is used as an encoder, (68) still holds.

With this encoder as (70), and $S$ be the confidential message $W$, $G = g$, (68) can be re-written as

$$I\left(W; t_1^{\bar{N}} \oplus t_2^{\bar{N}}, d, T\right) \leq 2e^{-\beta MN} \quad (71)$$

From Theorem 1, it can re-written as

$$I\left(W; X_1^N + X_2^N, d_1^N, d_2^N\right) \leq 2e^{-\beta MN} \quad (72)$$

Since $Y_2^N = X_1^N + X_2^N + Z_2^N$, from the data processing inequality, (72) implies

$$I\left(W; Y_2^N, d_1^N, d_2^N\right) \leq 2e^{-\beta MN} \quad (73)$$

Again the proof so far does not depend on the nature of $d_j^N, j = 1, 2$. As shown in [9], $d_j^N$ only affects the probability of decoding errors at the intended receiver and the average power of the transmitters, and we can choose $d_j^N$ as deterministic vectors. In this case, (73) is simply $I(W; Y_2^N)$. Hence we have proved the theorem.

## VII. PROOF OF THEOREM 3

We first introduce some useful results on extractor functions.

The definition of an extractor function can be found in [17, Definition 6]. For a random variable whose min-entropy can be lower bounded, [17] has the following lemma:

*Lemma 6:* [17, Lemma 9] Let $\delta', \Delta_1, \Delta_2 > 0$ be constants. Then there exists, for all sufficiently large $N$, an extractor function $E: \{0,1\}^N \times \{0,1\}^d \to \{0,1\}^r$, where $d \leq \Delta_1 N$ and $r \geq (\delta' - \Delta_2)N$, such that for all random variables $A$ whose alphabet is defined over $\{0,1\}^N$ and $H_\infty(A) > \delta' N$, we have

$$H(E(A,V)|V) \geq r - 2^{-N^{1/2} - o(1)} \quad (74)$$

The lemma says introducing a small amount of pure randomness, $V$, one can extract almost all min-entropy from a weakly random source. Its proof can be found in [17], which is based on an efficiently constructible extractor design from [21].

We use the same lattice codebook from [9] and follow the same notation in Section VI.

We begin with:

$$H_\infty\left(v(t_1^{\bar{N}})|t_1^{\bar{N}} \oplus t_2^{\bar{N}} = t^{\bar{N}}, d = \bar{d}\right) = H_\infty\left(v(t_1^{\bar{N}})\right) \quad (75)$$

According to Corollary 1, we have, for a given integer $a$, $1 \leq a \leq 2^{\bar{N}}$ and $t^{\bar{N}}$ taken from $\Lambda^M \cap \mathcal{V}(\Lambda_c)^M$, with probability $1 - 2^{-s}$:

$$H_\infty\left(v(t_1^{\bar{N}})|t_1^{\bar{N}} \oplus t_2^{\bar{N}} = t^{\bar{N}}, d = \bar{d}, T = a\right) \quad (76)$$

$$\geq H_\infty\left(v(t_1^{\bar{N}})|t_1^{\bar{N}} \oplus t_2^{\bar{N}} = t^{\bar{N}}, d = \bar{d}\right) - \log_2|T| - s \quad (77)$$

$$= \bar{N}_0 - \bar{N} - s \quad (78)$$

$$\geq \bar{N}(R_0 - 1) - s - 1 \quad (79)$$

where (79) follows from (56).

We then choose $r$ and $\delta'$ as in Lemma 6. This means for a $\Delta_2 > 0$, we choose $r \geq \bar{N}(R_0 - 1 - \Delta_2)$. Applying Lemma 6 we have the following bounds when (76)-(79) holds:

$$H\left(E\left(v\left(t_1^{\bar{N}}\right), V\right) | V, t_1^{\bar{N}} \oplus t_2^{\bar{N}} = t^{\bar{N}}, d = \bar{d}, T = a\right) \quad (80)$$

$$\geq r - 2^{-\bar{N}^{1/2} - o(1)} \quad (81)$$

Since the probability that (76)-(79) holds is at least $1 - 2^{-s}$, we have:

$$H\left(E\left(v\left(t_1^{\bar{N}}\right), V\right) | V, t_1^{\bar{N}} \oplus t_2^{\bar{N}}, d, T\right) \quad (82)$$

$$\geq \left(1 - 2^{-s}\right)\left(r - 2^{-\bar{N}^{1/2} - o(1)}\right) \quad (83)$$

Choose $s = \varepsilon \bar{N}$, where $0 < \varepsilon < R_0 - 1$. Then from (83), we observe that there exists $\beta > 0$, such that

$$I\left(E\left(v\left(t_1^{\bar{N}}\right), V\right); V, t_1^{\bar{N}} \oplus t_2^{\bar{N}}, d, T\right) \leq e^{-\beta \bar{N}^{1/2}} \quad (84)$$

The secret key is then generated from the following procedure: First node $S_1$ transmit the pure random sequence $V$ to $D_1$. According to Lemma 6, since the length of binary representation of $V$ is smaller than $\Delta_1 N$, where $\Delta_1$ can be arbitrarily small, the rate penalty of sending $V$ is negligible. Then node $S_1$ transmits the $t_1^{\bar{N}}$ over $N$ channel uses, while $S_2$ transmits $t_2^{\bar{N}}$, using the nested lattice coding scheme described in Section VI. $t_j^{\bar{N}}, j = 1, 2$ is chosen such that they are uniformly distributed over $\Lambda^M \cap \mathcal{V}(\Lambda_c)^M$ and $t_1^{\bar{N}}$ and $t_2^{\bar{N}}$ being independent. Finally node $D_1$ decode $t_1^{\bar{N}}$ using the algorithm given in [9]. The secret key is computed from $E\left(v\left(t_1^{\bar{N}}\right), V\right)$.

In this protocol, it is clear that the eavesdropper's knowledge is a degraded version of $V, t_1^{\bar{N}} \oplus t_2^{\bar{N}}, d, T$. From (84), we observe that $K$ is secure. Hence we have proved the theorem.

*Remark 4:* It is not clear how to invert the extractor function in [21] efficiently, which requires us to obtain $A$ from $E(A,V)$ and $V$. Because of this reason, the method is only used for secret key generation, but not for secret message transmission in this work.

## VIII. CONCLUSION

In this work, we developed coding schemes which provides strong secrecy by combining nested lattice codes with either universal hash function or the extractor function. In our previous work [8], the representation theorem for nested lattice codes is used to bound the Shannon entropy and prove weak secrecy rates. Here we showed the same theorem is also useful in bounding other information theoretic measure, i.e., the Rényi entropy and the min entropy, which in turn leads to strong secrecy results. With these coding schemes, we showed that for an interference channel with a confidential message, the same secrecy rate, and hence the same secure degree of freedom derived for weak secrecy is achievable for the strong secrecy notion as well. The rate region where both users have confidential messages to transmit can be obtained by time-sharing between the individual rates. Compared to previous strong secrecy scheme with nested lattice code, our scheme achieves faster decrease for the mutual information between the message and the eavesdropper's observation with respect to the number of channel uses.

## APPENDIX A
## PROOF OF LEMMA 3

Consider the set:

$$A = \left\{ t : \frac{\max_x \Pr(X = x | T = t)}{\max_x \Pr(X = x)} > 2^s \|T\| \right\} \quad (85)$$

Equation (12) is equivalent to $\Pr[t \in A] \leq 2^{-s}$. Suppose it is otherwise:

$$\Pr[t \in A] > 2^{-s} \quad (86)$$

Then we have

$$E_T\left[\max_x \Pr(X = x | T = t)\right] \quad (87)$$

$$= \sum_t \Pr(T = t) \max_x \Pr(X = x | T = t) \quad (88)$$

$$\geq \sum_{T \in A} \Pr(T = t) \max_x \Pr(X = x | T = t) \quad (89)$$

$$=2^s \|T\| \max_x \Pr(X=x) \sum_{T \in A} \Pr(T=t) \quad (90)$$

$$> \|T\| \max_x \Pr(X=x) \quad (91)$$

On the other hand, define $x_t^*$ as:

$$x_t^* = \arg\max_x \Pr(X=x|T=t) \quad (92)$$

Then we have:

$$\max_x \Pr(X=x) \quad (93)$$

$$\geq \Pr(X=x_t^*) \quad (94)$$

$$= \sum_t \Pr(X=x_t^*|T=t)\Pr(T=t) \quad (95)$$

$$\geq \Pr(X=x_t^*|T=t)\Pr(T=t) \quad (96)$$

$$= \max_x \Pr(X=x|T=t)\Pr(T=t) \quad (97)$$

Therefore

$$E_T\left[\max_x \Pr(X=x|T=t)\right] \quad (98)$$

$$= \sum_t \Pr(T=t)\max_x \Pr(X=x|T=t) \quad (99)$$

$$\leq \sum_t \max_x \Pr(X=x) \quad (100)$$

$$= \|T\| \max_x \Pr(X=x) \quad (101)$$

To obtain (100), we apply (93)-(97). (98)-(101) contradicts (87)-(91). Hence we have proved the lemma.